\documentclass[12pt]{article}
\usepackage[dvips]{graphics}
\usepackage{epsfig}

\usepackage{amsfonts}
\usepackage{amssymb}
\begin{document}
\thispagestyle{empty}

\hspace{8cm}{YerPhI Preprint-1624}
\begin{center}
N. M.~Agababyan $^{1)}$, L.~Grigorian$^{2)}$, N.~Grigoryan$^{2)}$,
H.~Gulkanyan $^{2)}$, \\ A. A.~Ivanilov $^{3)}$, Zh.~Karamyan
$^{2)}$, V.
A.~Korotkov$^{3)}$ 
\end{center}

\vskip 3.5cm
\begin{center}
{\normalsize \bf NUCLEAR ATTENUATION OF THREE-HADRON SYSTEMS \\
[.3cm] IN NEUTRINO-INDUCED REACTIONS}
\end{center}

\vskip 3.5cm
~\\[.5cm]
\rule {3.2cm}{1.pt}\\
$^{1)}$ Joint Institute for Nuclear Research, Dubna, Russia. \\
$^{2)}$ Alikhanian National Scientific Laboratory, Yerevan, Armenia.\\
$^{3)}$ Institute of High Energy Physics, Protvino, Russia. \\
\vskip 2.5cm
\begin{center}
Yerevan 2011
\end{center}

\baselineskip 22pt
\newpage
\thispagestyle{empty}

\begin{abstract}

For the first time, the nuclear attenuation of three hadron
systems is studied in neutrino-induced reactions using the data
obtained with SKAT bubble chamber. The strongest attenuation $(R_3
\sim 0.6)$ is observed for a system carrying an overwhelming
fraction of the current quark energy, as well as for a system with
the smallest effective mass. An indication is obtained that the
correlation effects in the nuclear attenuation play only a minor
role. The experimental data are compared with predictions of the
quark string fragmentation model.

\end{abstract}
\clearpage
\newpage
\pagenumbering{arabic}

\begin{center}
{\large 1. ~INTRODUCTION}\\
\end{center}

The investigation of the multihadron leptoproduction on nuclear
targets can provide a valuable information on the space-time
structure of the quark string fragmentation and the hadron
formation
([1--4] and references therein). At present the experimental data
on the double hadron production are available in electronuclear
[5] and neutrinonuclear [6] interactions. This work is devoted
to the study, for the first time, of the nuclear effects in the
neutrinoproduction of more complex, three hadron systems. In
Section 2 the experimental procedure is described. The
experimental data are presented and compared with theoretical
predictions in Section 3. The results are summarized in Section 4.

\begin{center}
{\large  2. ~EXPERIMENTAL PROCEDURE}\\
\end{center}

The experiment was performed with SKAT bubble [7] chamber exposed
to a wideband neutrino beam obtained with a 70 GeV primary protons
from the Serpukhov accelerator. The chamber was filled with a
propane-freon mixture containing 87 vol\% propane (C$_3$H$_8$) and
13 vol\% freon (CF$_3$Br) with the percentage of nuclei H:C:F:Br =
67.9:26.8:4.0:1.3\%. A 20 kG uniform magnetic field was provided
within the operating chamber volume.

Charged current interactions containing a negative muon with
momentum $p_{\mu} >$ 0.5 GeV/c were selected. Other negatively
charged particles were considered to be $\pi^-$ mesons. Protons
with momentum below 0.6 GeV$/c$ and a fraction of protons  with
momentum 0.6--0.85 GeV$/c$ were identified by their stopping in
the chamber. To non-identified positively charged particles the
pion mass was assigned. Events with errors in measuring the
momenta of all charged secondaries and photons less than 60\% and
100\%, respectively, were selected. Each event was given a weight
to correct for the fraction of events excluded due to improperly
reconstruction. More details concerning the experimental
procedure, in particular, the estimation of the neutrino energy
$E_{\nu}$ and the reconstruction of $\pi^0 \rightarrow 2 \gamma$
decays, can be found in our previous publications [8,9,10]. In
this work, $\gamma\gamma$ combinations with the effective mass
$m_{\gamma\gamma}$ closest to 135 MeV/$c^2$ (and enclosed in the
range of 105 $< m_{\gamma\gamma} <$ 165 MeV$/c^2$) were taken,
requiring the each detected $\gamma$ to be included in no more
than one $\gamma\gamma$ combination.

The events with $3 < E_{\nu} <$ 30 GeV were accepted provided that
the reconstructed mass $W$ of the hadronic system exceeds 2 GeV
and $y = \nu/E_\nu < 0.95$, $\nu$ being the energy transferred to
the hadronic system. No restriction was imposed on the transfer
momentum squared $Q^2$. The number of accepted events was 4499
(5683 weighted events). The contamination from the neutral current
(NC) interactions to the selected event sample was estimated to be
(2.6$\pm$0.2)\%. The mean values of the kinematic variables were
$\langle E_{\nu}\rangle$ = 10.0 GeV, $\langle W\rangle$ = 2.9 GeV,
$\langle W^2\rangle$ = 9.1 GeV$^2$, $\langle Q^2\rangle$ = 2.7
(GeV/c)$^2$ and $\langle \nu\rangle$ = 5.8 GeV.

The nuclear attenuation effects studied below are inferred with
the help of a comparison of the trihadron characteristics in two
event subsamples: the nuclear subsample $(B_A)$ and quasinucleon
subsample $(B_N)$ composed using a number of topological and
kinematical criteria \cite{ref8,ref9}. The $B_N$ subsample
includes events with no indication of the nuclear disintegration
or a secondary intranuclear interaction: the total charge of
secondary hadrons is required to be $q = +1$ (for the subsample
$B_n$ of interactions with a neutron) or $q = +2$ (for the
subsample $B_p$ of interactions with a proton), while the number
of recorded baryons (these included identified protons and
$\Lambda$ -hyperons, along with neutrons that suffered a secondary
interaction in the chamber) was forbidden to exceed unity, baryons
flying in the backward hemisphere being required to be absent
among them. Moreover, a constraint was imposed on the effective
target mass $M_t < 1.2$ GeV$/c^2$, the $M_t$ being defined as $M_t
= \Sigma(E_i - p_i^L)$ where the summation is performed over the
energies $E_i$ and the longitudinal momenta $p^L_i$ (along the
neutrino direction) of all recorded secondary particles. Events
that did not satisfy aforementioned criteria were included in the
subsample $B_S$ of cascade events. As a result, the weighted
numbers of the $B_p$, $B_n$ and $B_S$ proved to be 1249, 1304 and
3129.

The validity of the selection of the $B_p$ and $B_n$ subsamples
was verified \cite{ref9,ref11,ref12} by comparison of a large
number of the multiplicity and spectral characteristics of hadrons
in these subsamples with the available data obtained on hydrogen
and deuterium targets, resulting in a satisfactory agreement and
giving a sufficient ground to conclude that the $B_p$ and $B_n$
subsamples may contain only a minor contamination from events
where the secondary intranuclear interactions deteriorate the
characteristics of the primary hadrons. We checked the sensitivity
of the extracted data (presented in the next section) relative to
the choice of the boundary value of $M_t$, varying the latter from
1.1 to 1.4 GeV$/c^2$, e.g. changing the event numbers in the
$B_p$, $B_n$ and $B_S$ subsamples and continuously increasing the
contamination from secondary interactions in the subsamples $B_p$
and $B_n$ (see Fig. 1 in \cite{ref11}). We found that the
variation of the dihadron characteristics have been on an average
fivefold smaller as compared to the statistical errors.

As it follows from the composition of the propane-freon mixture
(see above) about 36\% of sub-sample $B_p$ is contributed by
interactions with free hydrogen (at the ratio of $\nu n$ and $\nu
p$ CC cross sections being equal to 2). Weighting the quasiproton
events with a factor of 0.64, one can compose a pure nuclear
subsample $B_A = B_S + B_n + 0.64 B_p$ and a quasinucleon
subsample $B_N = B_n + 0.64 B_p$. As it follows from the
percentage of C, F and Br nuclei in the propan-freon mixture the
mean atomic weight of the composite target is equal to
$\overline{A}$ = 27.6 for neutrinonuclear interactions. However,
since the nuclear attenuation effects are under consideration, it
seems more relevant to introduce an effective nucleus
$A_{\mathrm{eff}}$ in which the probability of the hadron
absorption is the same as that obtained from averaging over the
target nuclei. The calculations (see [13] for details) result in
19$\leq A_{\mathrm{eff}}\leq$23, depending on the momentum of the
neutrinoproduced hadron.

\begin{center}
{\large  3. ~THE RESULTS}\\
\end{center}

Below we will consider hadrons produced in the forward hemisphere
in the hadronic c.m.s. (i.e. in the region of $x_\mathrm F > 0$,
$x_\mathrm F$ being the Feynman variable), because in this region
the nuclear attenuation effects for hadrons dominate over
multiplication effects caused by secondary intranuclear inelastic
interactions, observable in the region of $x_\mathrm F < 0$ and,
to a much smaller extent, even in the region of $0< x_\mathrm F <
0.1$ [9,11]. Below we will imply two different cuts on $x_\mathrm
F$: $x_\mathrm F > 0$ and $x_\mathrm F > 0.1$.

For a given kinematic variable $v$, characterizing a trihadron,
the nuclear attenuation is measured by the ratio $R_3(v) =
n_3^A(v)/n_3^N(v)$, where $n_3^A(v)$ and $n_3^N(v)$ are the
trihadron differential mean multiplicities (yields) in the nuclear
and quasinucleon subsamples, respectively.

Figure 1 (top panel) shows the dependence of $R_3$ on the maximal
angle $\vartheta^\mathrm{max}$ between two hadrons entering into
the three-hadron system. The data exhibit a strengthening
attenuation with decreasing $\vartheta^\mathrm{max}$ which can be
caused by secondary intranuclear interactions, mainly elastic for
the case of $x_\mathrm F > 0.1$ and both elastic and inelastic for
the case of $x_\mathrm F > 0$. The elastic scattering of hadrons
also leads to the increasing effective mass $m_\mathrm{eff}$ of
the trihadron and hence to a stronger attenuation at smaller
$m_\mathrm{eff}$ than at larger $m_\mathrm{eff}$, as it is seen
from the middle panel of Figure 1 for $x_\mathrm F > 0.1$. On the
other hand, the inelastic interactions (playing more significant
role at the cut $x_\mathrm F > 0$) lead to energy losses of
hadrons which partly compensate the effective mass increasing
caused by the widening of the angles between scattered hadrons, as
it can be concluded from the middle panel of Figure 1 where a
rather flat $m_\mathrm{eff}$ - dependence is observed for the case
of $x_\mathrm F > 0$.

The bottom panel of Figure 1 shows the dependence of the ratio
$R_3(p_t)$ on the trihadron transverse momentum $p_t$ (defined
relative to the current quark direction). A trend of increasing
$R_3(p_t)$ with increasing $p_t$ is observed, being a consequence
of secondary interactions of neutrinoproduced hadrons. It should
be noted, that a characteristic feature of the data presented in
Figure 1 is that, in general, the ratio $R_3(v)$ at $x_\mathrm F
> 0$ exceeds that for the case of $x_\mathrm F > 0.1$, indicating, as
it was already mentioned above, that a non-negligible fraction of
low $x_\mathrm F$ hadrons (with $0 < x_\mathrm F < 0.1$)
originates from inelastic intranuclear interactions of higher -
$x_\mathrm F$ hadrons.

Figure 2 presents the dependence of the ratio $R_3$ on the
collective variable $Z = \sum{z_k}$, where $z_k = E_k/\nu$ with
$E_k$ being the energy of the $k$- th hadron and the sum is over
three hadrons. For comparison, we also plotted the ratio $R_2(z)$
versus $Z=z_1+z_2$ for two-hadron systems and the ratio $R_1(z)$
versus the $z$ value for single hadrons. As it is seen, both
functions $R_1$, $R_2$ and $R_3$ continuously decrease with the
increasing argument. It worths to note, that the fastest single
hadron (with $z > 0.85$) absorbs stronger than the fastest hadron
systems (with the same summary $Z$) composed of two or three
hadrons of smaller $z$'s.

It is also interesting to compare the ratio $R_1(z)$ for single
hadron with the ratios $R_2(z)$ and $R_3(z)$ where the argument
refers to a hadron (the "trigger" hadron) accompanied by,
respectively, one and two hadrons of arbitrary values of the
variable $z$. A decreasing $n$- dependence of $R_n(z)$, seen in
Figure 3 and, more clearly, in Figure 4 and 5, is related to the
additional attenuation of the accompanying hadron(s). The $n$-
dependencies of $R_n(z)$ plotted in Figure 4 (at $x_\mathrm F >
0$) and 5 (at $x_\mathrm F > 0.1$) can be approximated as $R_n(z)
= \rho_1(z)(1-w_1)^{n-1}$, where $\rho_1(z)$ is the attenuation
ratio for the 'trigger' hadron, while $w_1$ is the mean absorption
probability of a single hadron averaged over hadrons accompanying
the 'trigger' one. As it is seen from the fitted values of
$\rho_1(z)$ and $w_1$ quoted in Figures 4 and 5, the ratio
$\rho_1(z)$ is maximal for the slowest trigger hadron (even
exceeding unity at $x_\mathrm F > 0$ due to the nuclear
enhancement effects). At higher $z$- values, the variation of
$\rho_1(z)$ is not significant, while the values of $w_1$ are
around $w_1 \sim 0.05$ at $x_\mathrm F > 0$ and around $w_1 \sim
0.12$ at $x_\mathrm F > 0.1$. These $w_1$ values are averaged over
a wide range of the $z$ variable of accompanying hadrons. A more
definite estimation for the absorption probability can be inferred
from the $n$- dependence of $R_n(z)$ at a narrow $z$- range of all
involved hadrons, for example, at $0.1 < z < 0.33$ (see Figure 6).
Under assumption that hadrons absorb independently with a mean
probability $w_\mathrm{abs}$, the $n$- dependence of $R_n(z)$ can
be expressed as $R_n(z) = (1 - w_\mathrm{abs})^n$. The data at
$x_\mathrm F > 0$ are described by this dependence quit well (with
the fit value $w_\mathrm{abs} = 0.060\pm0.006$), being consistent
with the assumption that the correlated absorption effects do not
play a prominent role (Figure 6, the left panel). The description
of the data at $x_\mathrm F > 0.1$ (with the fit value
$w_\mathrm{abs} = 0.077\pm0.014$) is somewhat worse, due to a
possible (statistically not provided) deviation of the $R_3$ value
from the general trend of the $n$- dependence of $R_n$. It can be
a faint manifestation of the mutual screening effect (suggested in
[4]) in the system of hadrons produced at close impact parameters.
Statistically more provided data are needed to check this
assumption.

It should be also noted, that the observed difference between
$w_1$ and $w_\mathrm {abs}$ (quoted in Figures 5 and 6 for the
case of $x_\mathrm F > 0.1$) can be caused by the fact that the
latter concerns the hadrons with $0.1 < z < 0.33$ at which the
hadron formation length is expected to be maximal (hence reducing
their absorption), while the former concerns the hadrons which can
carry larger $z$'s at which the formation length is comparatively
shorter [14].

We undertook an attempt to describe our data in the framework of
the Two-Scale Model (TSM) [15] (see also [4] and references
therein) which was already applied for description of our data on
the nuclear attenuation of neutrinoproduced two-hadron systems
[6]. The details of the TSM application to three-hadron systems
can be found in [16]. The TSM contains four free parameters: the
quark string tension $\kappa$ and the following string-nucleon
interaction cross sections, namely, $\sigma_q$ for the initial
string stretched between the struck quark and the target nucleon
remnant, $\sigma_s$ for the intermediate string stretched between
the struck quark and a created antiquark (which becomes a valence
one for the hadron being looked at) and $\sigma_h$ for the formed
colorless system with quantum number and valence content of the
final hadron.

As it was shown in [6], a reasonable agreement with the data on
the two-hadron neutrinoproduction can be reached at the following
set of the model parameters: $\kappa$ = 0.8 GeV/fm. $\sigma_q$ =
0, $\sigma_s$ = 10 mb and $\sigma_h$ = 20 mb. This set was used to
get the TSM predictions presented below. The predictions concern
pions, pion pairs and pion triplets. In present calculations, only
the $z$ variable of produced pions is considered, while the $x_F$
variable does not appear explicitly in the model. Therefore, the
same predictions of the model are compared below, in Figs. 2, 3,
4, 5 and 6, with experimental data inferred both at $x_F > 0$ and
$x_F > 0.1$.

As it is seen, the description of the data at $x_F > 0$ is
noticeably worse, probably, due to the aforementioned effects of
intranuclear interactions (not inserted into the model) leading to
a significant enhancement of the yield of low-$z$ hadrons
(especially in the region of $z <0.2$) and, as a consequence, the
yields of dihadrons and trihadrons with comparatively small values
of $Z< 0.3 \div 0.4$. At a more severe restriction on $x_F > 0.1$,
the said discrepancy is much smaller, and the model provides a
reasonable description of the data at $Z > 0.2$, as it can be seen
from Figs. 2 and 3. An exception is $R_1(z)$ for the fastest
single hadron with $z > 0.85$ for which the model badly
underestimates the nuclear attenuation. On the other hand, the
model predictions are compatible with the data at the the same
summary $Z$-range ($Z \sim 0.9$) for hadron systems composed of
two or three hadrons with smaller individual $z$'s (see Fig. 2).
The TSM describes quite well the $n$-dependence of the ratio
$R_n(z)$ at $z$-values of the trigger hadron above $z$ = 0.22, as
it can be seen from Fig. 3 (the right panel) and Fig. 5 (the
middle and right panels). It should be also noted, that the TSM
predictions for the slope in the $n$-dependence of $R_(z)$ (the
dashed lines in Fig. 5) at $z > 0.22$ are rather close to those
extracted under assumption of an independent absorption of
neutrinoproduced hadrons (the solid lines in Fig. 5). The same can
be said about the slopes in the $n$-dependence of $R_n$ for the
case when all involved hadrons are enclosed in a narrow $z$-range
of $0.1 < z < 0.33$, as it is seen from Fig. 6 where the model
predictions are plotted by dashed lines. The predicted values of
$R_n$, however, underestimate the experimental data, probably, due
to the unsufficiently large values of the variable $z$ (see the
discussion above).

\begin{center}
{\large  4. ~SUMMARY}\\
\end{center}

The nuclear attenuation of three-charged hadron systems is studied
for the first time in neutrino-induced reactions.

The dependence of the attenuation factor $R_3$ on various
kinematic variables of the trihadron is investigated. The
strongest attenuation, with $R_3 \sim 0.6$, is observed for a
system of hadrons with $x_\mathrm F > 0.1$ having the smallest
effective mass, as well as at largest values of the collective
variable $Z$. The observed dependence of $R_n$ on the number $n$
of involved hadrons enclosed in the narrow range of $0.1 < z <
0.33$ is compatible with an assumption that the intranuclear
interaction of a hadron occurs with a mean probability
$w_\mathrm{abs} = 0.07\pm0.02$ and almost independently of
accompanying hadrons.

It is shown, that the Two-Scale Model of the quark string
fragmentation provides a reasonable description of the nuclear
attenuation for systems composed of relatively energetic ($x_F >
0.1$ and $z > 0.2$) hadrons.

\newpage

\begin{center}
{\large ACKNOWLEDGMENTS}\\
\end{center}

The authors from YerPhI acknowledge the
supporting grants of Calouste Gulbenkian Foundation and Swiss
Fonds Kidagan. The activity of two of the authors (L.G. and
H.G.) is supported by Cooperation Agreement between DESY and
YerPhI signed on December 6, 2002.


\begin{center}

\end{center}
%


\newpage
\begin{figure}[ht]
\resizebox{0.9\textwidth}{!}{\includegraphics*[bb =20 60 600
610]{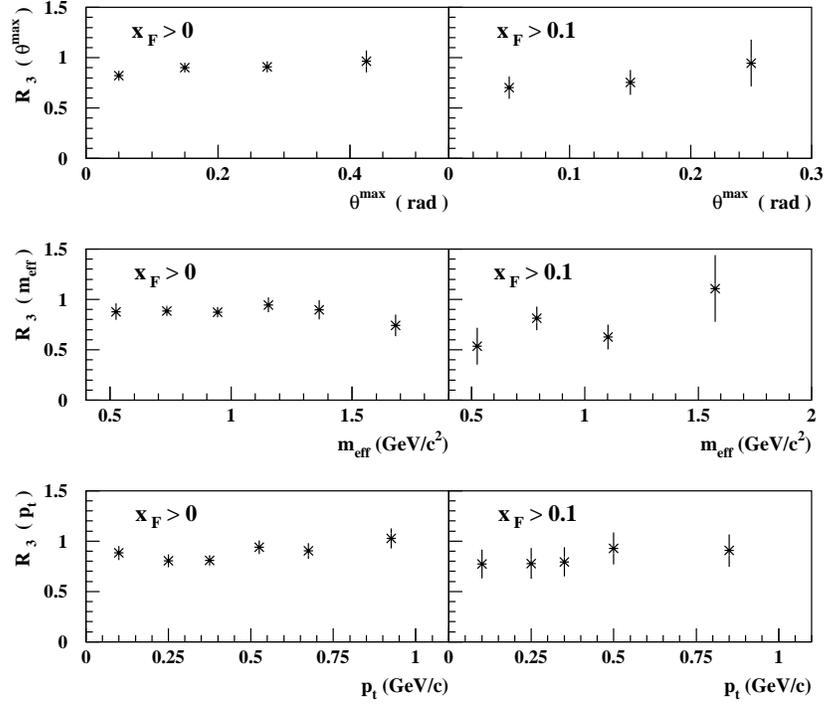}} \caption{The dependence of $R_3$ on
$\vartheta^\mathrm {max}$ (top panels), $m_\mathrm {eff}$ (middle
panels) and $p_t$ (bottom panels) for hadrons with $x_\mathrm F >
0$ (left panels) and $x_\mathrm F > 0.1$ (right panels).}
\end{figure}

\newpage
\begin{figure}[ht]
\resizebox{0.9\textwidth}{!}{\includegraphics*[bb =20 60 600
610]{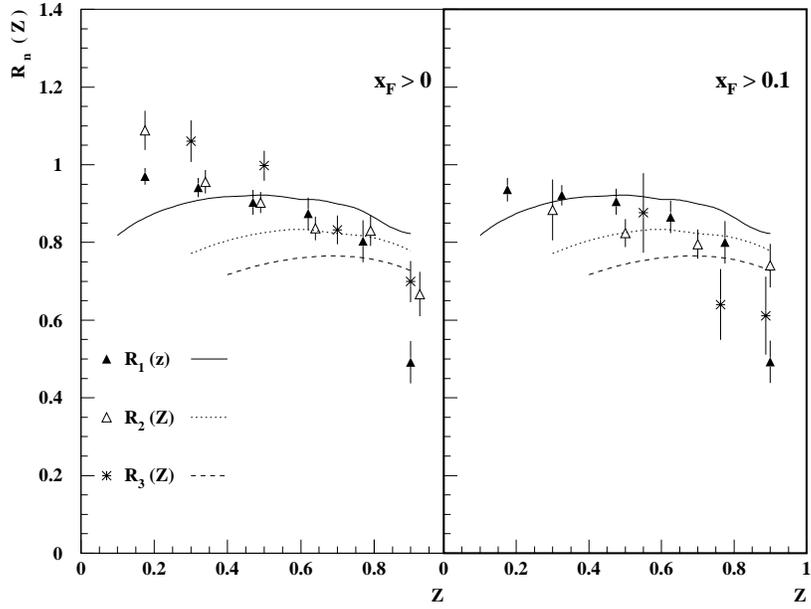}} \caption{The dependence of $R_n$ ($n = 1,2,3)$ on
$Z$ for hadrons with $x_\mathrm F
> 0$ (left panel) and $x_\mathrm F > 0.1$ (right panel) The curves are the TSM
predictions (see text).}
\end{figure}

\newpage
\begin{figure}[ht]
\resizebox{0.9\textwidth}{!}{\includegraphics*[bb =20 60 600
610]{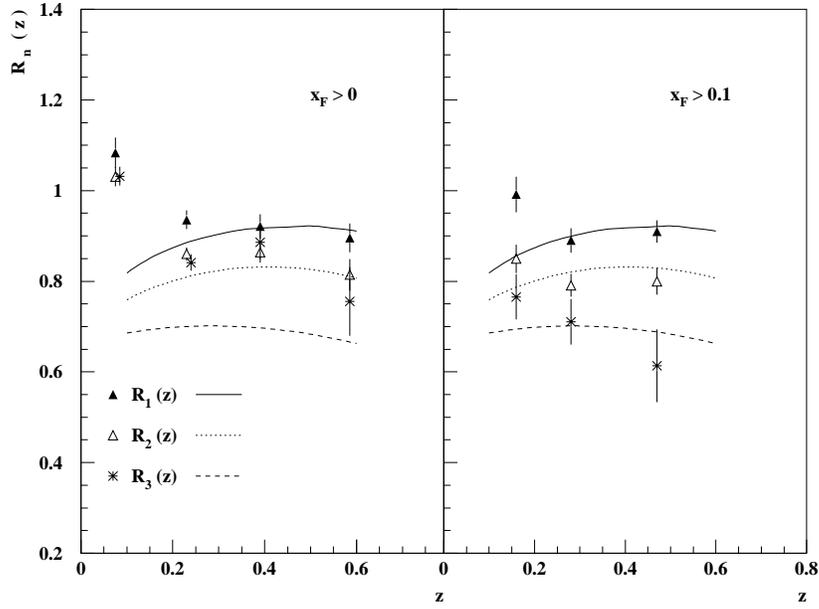}} \caption{The dependence of $R_n(z)$ on the $z$
variable of the "trigger" hadron at $x_\mathrm F > 0$ (the left
panel) and $x_\mathrm F > 0.1$ (the right panel) The curves are
the TSM predictions (see text).}
\end{figure}

\newpage
\begin{figure}[ht]
\resizebox{0.9\textwidth}{!}{\includegraphics*[bb =20 60 600
610]{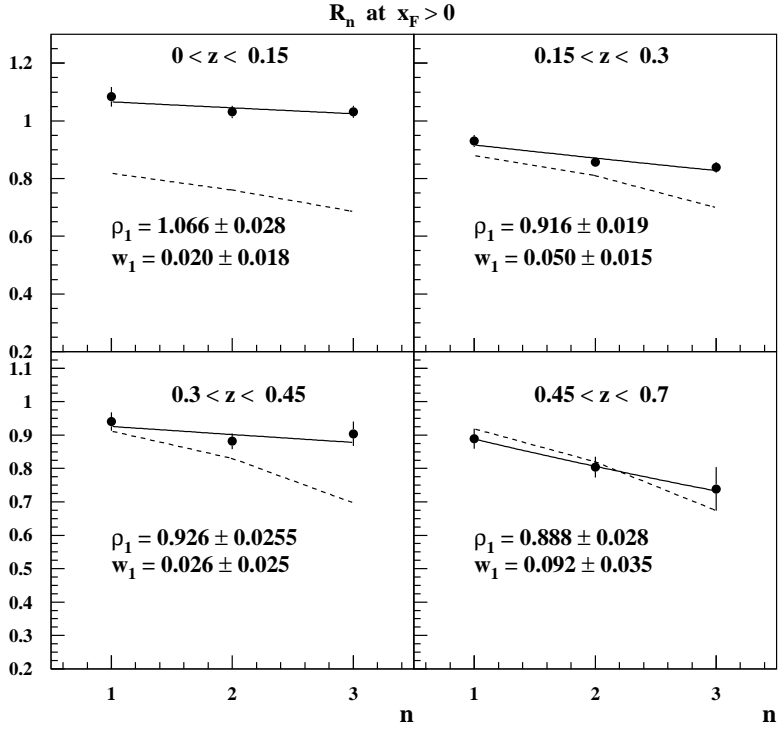}} \caption{The $n$- dependence of $R_n(z)$ for
different $z$- ranges of the "trigger" hadron at $x_\mathrm F
> 0$. The solid lines are the fit results, while the dashed lines
are the TSM predictions (see the text).}
\end{figure}

\newpage
\begin{figure}[ht]
\resizebox{0.9\textwidth}{!}{\includegraphics*[bb =20 60 600
610]{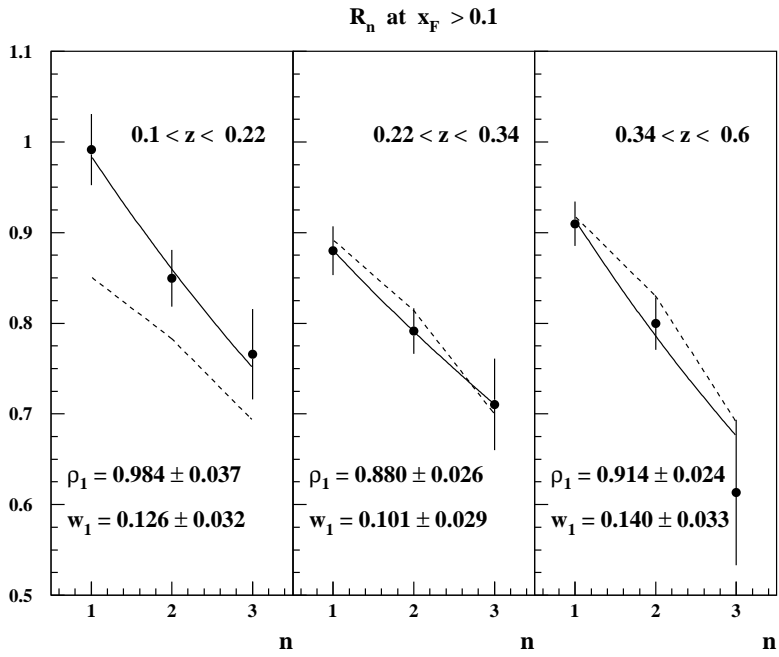}} \caption{The same as Fig. 4, but at $x_\mathrm F
> 0.1$.}
\end{figure}

\newpage
\begin{figure}[ht]
\resizebox{0.9\textwidth}{!}{\includegraphics*[bb =20 60 600
610]{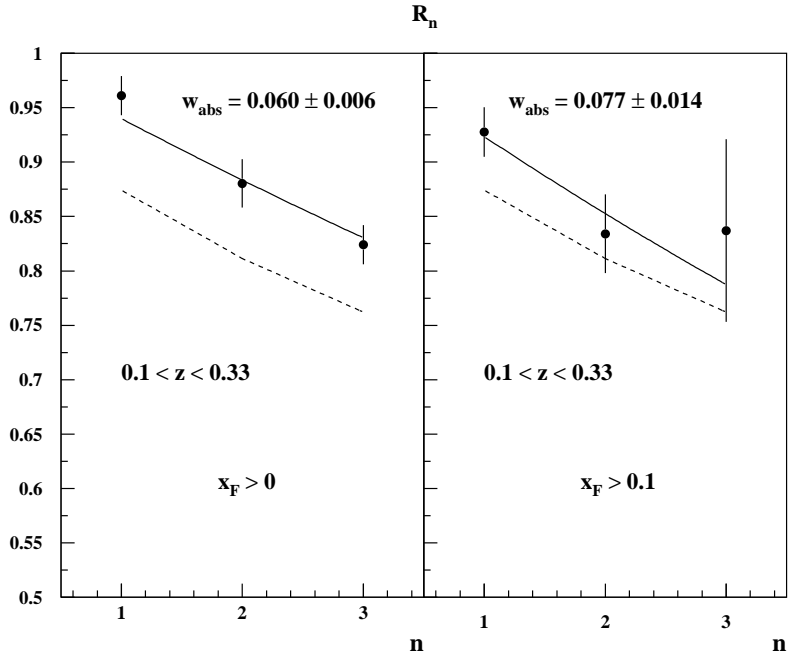}} \caption{The dependence of $R_n(z)$ on the number
$n$ of involved hadrons with $0.1 < z < 0.33$ and $x_\mathrm F
> 0$ (the left panel) and $x_\mathrm F
> 0.1$ (the right panel). The solid lines are the fit results, while the
dashed lines are the TSM predictions (see the text).}
\end{figure}

\end{document}